\documentclass[%
 aip,
 amsmath,amssymb,
 reprint,%
]{revtex4-1}

\usepackage{graphicx}
\usepackage{dcolumn}
\usepackage{bm}

\usepackage[utf8]{inputenc}
\usepackage[T1]{fontenc}
\usepackage{mathptmx}
\usepackage{etoolbox}
\usepackage{xcolor}
\usepackage{makecell}
\usepackage{xr}
\makeatletter
\def\@email#1#2{%
 \endgroup
 \patchcmd{\titleblock@produce}
  {\frontmatter@RRAPformat}
  {\frontmatter@RRAPformat{\produce@RRAP{*#1\href{mailto:#2}{#2}}}\frontmatter@RRAPformat}
  {}{}
}%
\makeatother

\begin{document}

\preprint{AIP/123-QED}

\title{Molecular beam epitaxy growth of superconducting tantalum germanide}
\author{Patrick J. Strohbeen}
\affiliation{ 
Center for Quantum Information Physics, Department of Physics, New York University, New York, NY 10003 USA
}%
\author{Tathagata Banerjee}
\affiliation{
School of Applied and Engineering Physics, Cornell University, Ithaca, NY 14853 USA
}%
\author{Aurelia M. Brook}
\author{Ido Levy}
\affiliation{ 
Center for Quantum Information Physics, Department of Physics, New York University, New York, NY 10003 USA
}%
\author{Wendy L. Sarney}
\affiliation{%
Army Research Directorate, DEVCOM Army Research Laboratory, Adelphi, MD 20783 USA
}%
\author{Jechiel van Dijk}
\author{Hayden Orth}
\author{Melissa Mikalsen}
\author{Valla Fatemi}
\affiliation{
School of Applied and Engineering Physics, Cornell University, Ithaca, NY 14853 USA
}%
\author{Javad Shabani}
\thanks{corresponding author: jshabani@nyu.edu}
\affiliation{ 
Center for Quantum Information Physics, Department of Physics, New York University, New York, NY 10003 USA
}%

\date{\today}

\begin{abstract}
Developing new material platforms for use in superconductor-semiconductor hybrid structures is desirable due to limitations caused by intrinsic microwave losses present in commonly used III/V material systems. With the recent reports on tantalum superconducting qubits that show improvements over the Nb and Al counterparts, exploring Ta as an alternative superconductor in hybrid material systems is promising. Here, we study the growth of Ta on semiconducting Ge (001) substrates grown via molecular beam epitaxy. We show that at a growth temperature of 400$^{\circ}$C the Ta diffuses into the Ge matrix in a self-limiting nature resulting in smooth and abrupt surfaces and interfaces with roughness on the order of 3-7 \r{A} as measured by atomic force microscopy and x-ray reflectivity. The films are found to be a mixture of Ta$_{5}$Ge$_{3}$ and TaGe$_{2}$ binary alloys and form a native oxide that seems to form a sharp interface with the underlying film. These films are superconducting with a $T_{C}\sim 1.8-2$K and $H_{C}^{\perp} \sim 1.88T$, $H_{C}^{\parallel} \sim 5.1T$. These results show this tantalum germanide film to be promising for future superconducting quantum information platforms.
\end{abstract}

\maketitle

Superconductor-semiconductor (S-Sm) hybrid material platforms have been of interest in the last few decades for studying mesoscopic superconductor physics \cite{rahman1996ssmmeso}, the search for topological superconductivity \cite{mourik2012majorana, prada2020absmajreview}, and most recently for the development of new voltage-tunable qubits, couplers, and other superconducting circuit elements \cite{casparis2018inasgatemon, burkard2020supersemicqed, strickland2022tuneresi}. However, due to the complexity of the material requirements (e.g. interface roughness, band alignment, etc.), growth of these structures is not straightforward. Furthermore, in terms of the latter application in superconducting circuitry for circuit quantum electrodynamics (cQED) applications, high intrinsic losses in the community standard Al-InAs system \cite{casparis2018inasgatemon} have motivated material exploration studies to search for materials better suited towards these cQED applications. On the other hand, recent advancements in the Si-Ge alloy system present these materials to be highly promising for low-microwave loss materials \cite{sandberg2021sigeloss, scappucci2021geinfo, lodari2022strainge, sarkar2023elecopge, myronov2023holesoutperform, hartmann2023grpivepi, myronov2023zeemansige, li2023sigetransport, tosato2023gehardgap, kong2023ge2mil} for quantum information applications.

In terms of superconducting materials, implementation of tantalum metal is also extremely interesting for cQED devices due to the fact that it forms a well-behaved native oxide for superconducting microwave applications \cite{douglass1972supercondmetals, spencer1981tazrsupercond, face1986tamicrowavemix}. Additionally, recent studies showed that this behavior of Ta also propagates to high-performance cQED devices \cite{place2021taqubit, mclellan2023taoxidexps}. In this context, major increases in qubit coherence times are caused by low intrinsic two-level system (TLS) losses within the Ta metal native oxide, Ta$_{2}$O$_{5}$ \cite{place2021taqubit, mclellan2023taoxidexps}. However, implementation of this material into S-Sm hybrid materials requires significant effort in terms of understanding the film growth parameters to promote smooth interfaces/surfaces. Thus, further study of Ta and Ta-alloys is of interest for investigation of new superconductor-semiconductor (S-Sm) hybrid material systems.

Here, we report the diffusion-limited growth (diffusion-growth) of tantalum germanide films via molecular beam epitaxy (MBE). This technique is similar to what has been previously employed to study superconducting contacts in GaAs/AlGaAs systems \cite{gao1996snticontacts}. We evaluate our grown films via a suite of structural and electronic materials characterization and present the chemical composition, structure, and superconductivity behavior of this material. We demonstrate a unique growth method that forms uniform, wafer-scale thin films of homogeneous chemical composition and thickness that make this material system highly promising for future application in superconducting cQED devices.

The tantalum germanide films are grown on 50.8mm Ge(001) substrates in a custom solid-source molecular beam epitaxy system (Mod Gen II) equipped with three electron beam sources and hydrogen cleaning capabilities. The internal manifold around the sources is water-cooled and no L$N_{2}$ cryo-paneling is used during this deposition. The substrates are first etched \textit{ex-situ} in deionized water at 90$^{\circ}$C for 15 minutes and then immediately loaded into vacuum on indium-free blocks. The substrates are initially outgased in the growth reactor at 200$^{\circ}$C for 15 minutes, before slowly increasing temperature to 550$^{\circ}$C to anneal for 10 minutes. We confirm the removal of the surface oxide via \textit{in-situ} reflection high-energy electron diffraction (RHEED) monitoring. We use the presence of a well-defined and sharp (2x1) surface reconstruction of the Ge (001) facet to indicate the removal of any native oxide the reforms during the substrate mounting process. Ta deposition was done using a water-cooled vertical EBVV e-beam source from MBE Komponenten at an emission current of 180mA in a 6kV acceleration field. The films are grown at 400$^{\circ}$C, immediately following the Ge oxide removal. Due to passive heating from the Ta source, deposition was conducted in three cycles of 10 minutes each, allowing the chamber to cool for 15 minutes between deposition cycles to prevent significant outgassing from the chamber walls. 

\begin{figure*}[h!]
    \centering
    \includegraphics[width=\linewidth]{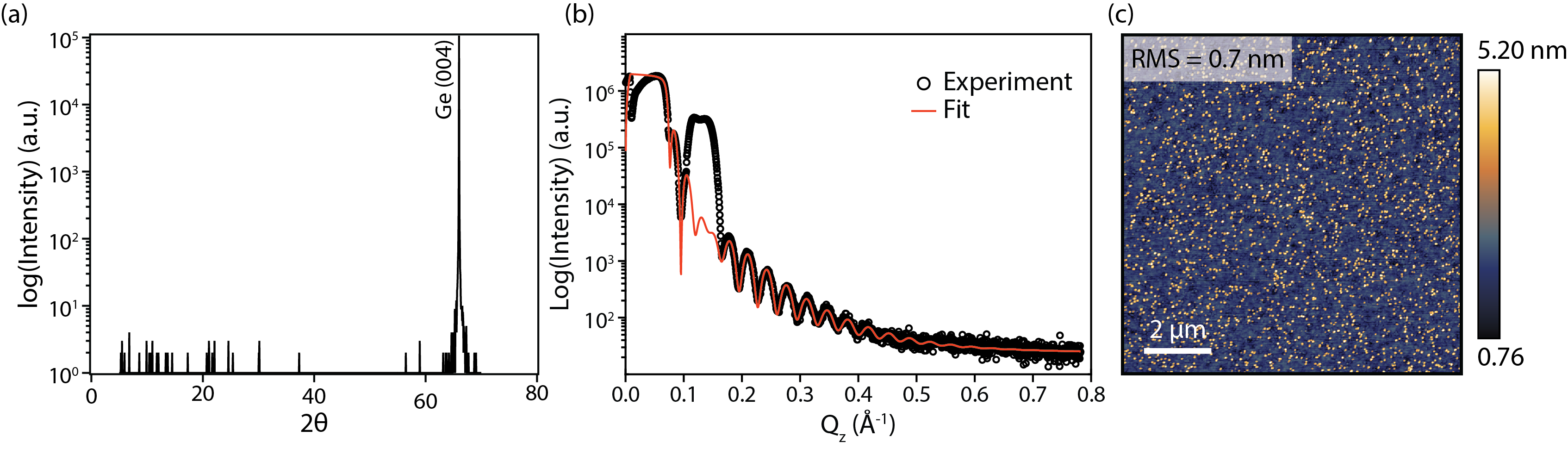}
    \caption{X-ray diffraction characterization of tantalum germanide film. \textbf{(a)} Out-of-plane $\theta$-$2\theta$ scan. The only distinct reflection is from the Ge substrate. \textbf{(b)} X-ray reflectivity measurement of the tantalum germanide film. The fit is done using REFLEX \cite{vignaud2019reflexxrr} standalone reflectivity fitting software, assuming two layers of composition TaGe$_{2}$ and Ta$_{2}$Ge$_{2}$O$_{9}$ as initial conditions. \textbf{(c)} 10$\mu m$x10$\mu m$ AFM image of tantalum germanide surface. RMS value of 7 $\mathring{A}$ was used as the initial roughness for the XRR fitting.}
    \label{xrd_structure}
\end{figure*}

Figure \ref{xrd_structure} presents x-ray diffraction data taken using a Bruker D8 Discovery lab source diffractometer with a da Vinci configuration and a conditioned Cu K$\alpha_{1}$ source. A collimator and 1mm slit are used to reduce the effects of substrate bowing in select measurements. All scans are measured in a double crystal configuration. An out-of-plane $\theta-2\theta$ scan is presented in Fig. \ref{xrd_structure}a in which the only reflection visible is the substrate Ge (004) reflection at 66deg. X-ray reflectivity (XRR) measurements are taken after removing the collimator and reducing the slits to 0.2mm. The XRR results are presented in Figure \ref{xrd_structure}b, showing extremely sharp interfaces for the amorphous film. The fit is initialized assuming a two-layer model for the film structure with the primary layer being a film of tantalum germanide, composition TaGe$_{2}$, and the secondary layer being a thin native oxide layer of composition Ta$_{2}$Ge$_{2}$O$_{9}$. Since the density is unknown due to the amorphous nature, we allow the scattering length density (SLD) to vary in the fitting procedure. The results of the fit are shown below in table \ref{xrr_table}. The feature that is not captured by our fitting model we speculate is related to the dot-like features observed in the atomic force microscopy (AFM) image seen in Figure \ref{xrd_structure}c. These features are well dispersed across the sample surface and are all nominally a uniform thickness and may be the cause of this spurious peak that is not captured by our two-layer model. The roughness values for the fits are initialized at 7 $\mathring{A}$, the roughness we measured in AFM, but then is allowed to vary for both layers individually.

\begin{table}[h!]
    \centering
    \begin{tabular}{||c|c|c|c||}
        \hline
        Layer & \thead{SLD \\ ($10^{-6}/\mathring{A}^{2}$)} & \thead{thickness \\ ($\mathring{A}$)} & \thead{Roughness \\ ($\mathring{A}$)} \\
        \hline\hline
        \makecell{Tantalum \\ Germanide} & 93.97 & 175.1 & 3.6 \\
        \hline
        \makecell{Native \\ Oxide} & 64.95 & 24.4 & 9.3 \\ 
        \hline
    \end{tabular}
    \caption{X-ray reflectivity fitting results.}
    \label{xrr_table}
\end{table}

From the fits we extract scattering length densities (SLD) for total internal reflection x-ray scattering in the XRR measurements. Using Eqn. \ref{sld_eq} below we then calculate the product of $\rho f_{1}/M_{a}$, relating the SLD from the fit back to the material density and scattering factor. 

\begin{equation}
    {\rm Re}(SLD) = \frac{\rho N_{a}r_{e}}{M_{a}}f_{1}
    \label{sld_eq}
\end{equation}

Where $\rho$ is the material density, $N_{a}$ is Avogadro's constant, $r_{e}$ is the classical electron radius $\sim 2.818 \times 10^{-15}$~m, $M_{a}$ is the molar mass, and $f_{1}$ is the real part of the atomic scattering factor of the compound. We calculate values of $5.54 cm^{-3}$ and $3.83 cm^{-3}$ for the tantalum germanide and native oxide layers, respectively.

\begin{figure}[h!]
    \centering
    \includegraphics[width=0.8\linewidth]{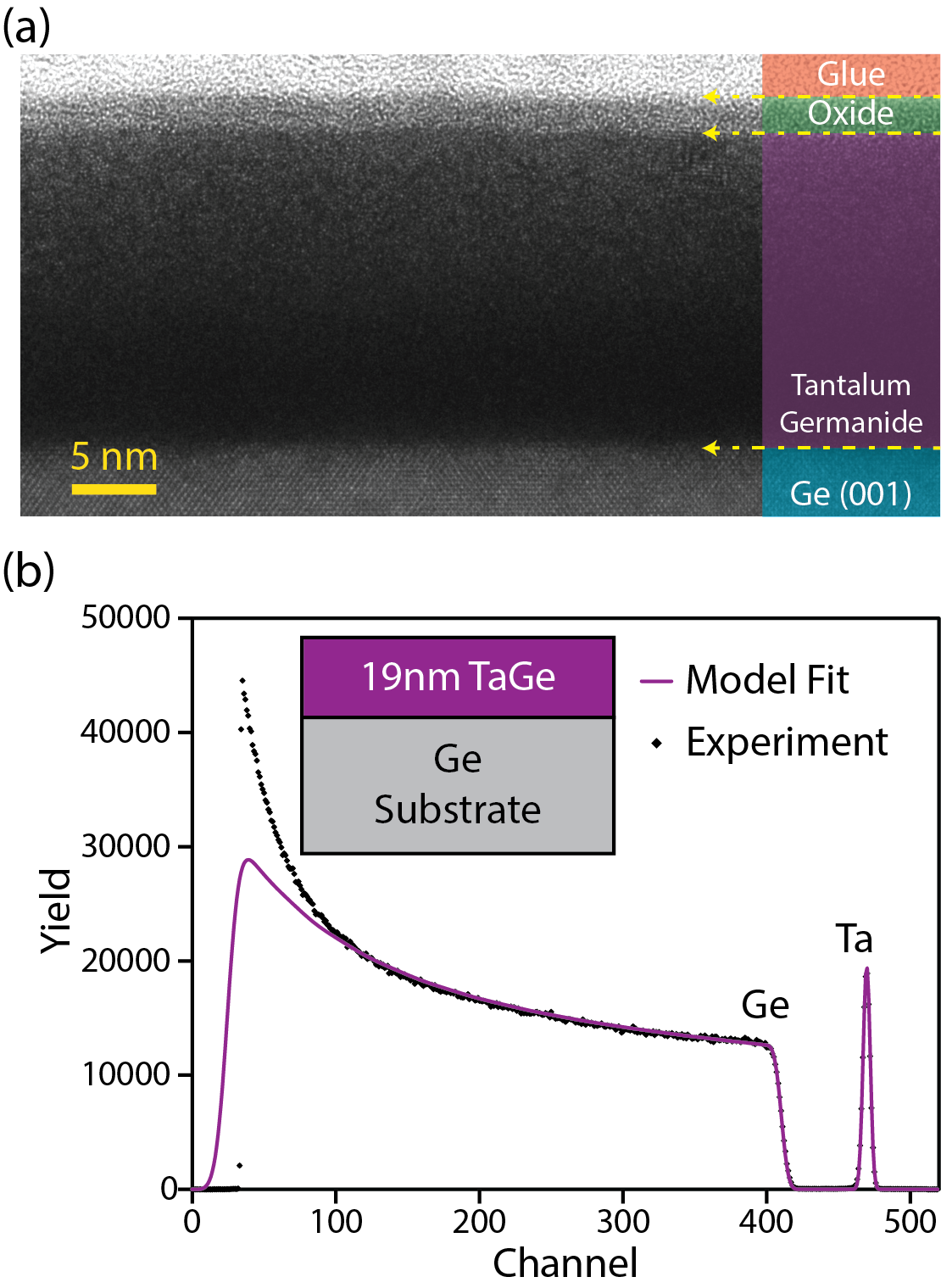}
    \caption{TEM imaging of tantalum germanide film. \textbf{(a)} Zoomed-out image of tantalum germanide lamella that shows sharp interfaces at the oxide-film and film-substrate interfaces. \textbf{(b)} RBS trace for the tantalum germanide film. The black dots and trace are the experimental data and the purple line is the model fit. The inset presents the model used to generate the fit.}
    \label{TEM_structure}
\end{figure}

Crystallinity and interface structure of the tantalum germanide film is examined with scanning transmission electron microscopy (STEM) in a JEOL ARM200F, equipped with a spherical aberration corrector for probe mode, and operated at 200 keV. The samples were prepared with cross-sectional tripod polishing to 20 $\mu$m thickness, followed by shallow angle Ar+ ion milling with low beam energies([1]3 keV), and LN2 stage cooling in a PIPS II ion mill. Exemplary transmission electron microscopy images are shown in Figure \ref{TEM_structure}, where we observe sharp interfaces and well-defined film and oxide regions of thicknesses $\sim$18.7~nm and $\sim$2.2~nm for the film and oxide, respectively, which are in good agreement with the XRR measurements. Energy dispersive X-ray spectroscopy (EDS) was conducted in an Oxford Aztec system and is presented in Fig. \ref{s1-eds} in the Supplemental material. From compositional mapping, a thickness of $\sim$18.2~nm for the film is found, in good agreement with our TEM imaging.

For an accurate determination of the full film composition, Rutherford Backscattering Spectroscopy measurements were conducted via EAG Eurofins, using a Cornell measurement geometry. The He$^{++}$ ion beam energy was set to 2.275~MeV He$^{++}$, and the detector was placed at an angle of 160$^{\circ}$ from the ion beam path. A second grazing incidence detector was placed at 100$^{\circ}$ to increase measurement accuracy. Fig. \ref{TEM_structure}b presents the RBS data collected at the detector 160$^{\circ}$ from the beam path. From these measurements we calculate the film stoichiometry to be $\sim 58.5\pm0.5:41.5\pm1$ Ta:Ge ratio, or a chemical formula of roughly Ta$_{3}$Ge$_{2}$. Using the Ta-Ge binary phase diagram \cite{dasilva2013tagephase}, the film is likely a solid solution of 86.3\% Ta$_{5}$Ge$_{3}$ and 13.7\% TaGe$_{2}$. Exact composition of the oxide layer is unable to be determined via RBS. Depth-resolved X-ray photoemission spectroscopy (XPS) shows a nominally constant composition throughout the film layer, as seen in the Supplementary Materials in Fig. \ref{s2-xps}, similar to what is observed in the EDS measurement. The composition of the film as measured in XPS is more consistent with the RBS measurement and the oxide region is similar to what is observed in EDS. We note that the three different methods of RBS, XPS, and EDS report varying Ta concentrations where XPS reports the film to be Ta-rich, EDS reports a stark Ta deficiency, and RBS reports a Ta$_{5}$Ge$_{3}$-rich mixture of Ta$_{5}$Ge$_{3}$ and TaGe$_{2}$ phases. For the remainder of this work, we take the RBS measurement to be the true absolute measure of film composition.

For determining the density of the film, we use the value of composition as reported from our RBS measurements. Assuming random distribution of elemental species in the two distinct layers, we then calculate the expected average material density to be $14.39 g/cm^{3}$ for the tantalum germanide layers, respectively. Comparing against the density of the crystalline form, approximated to be $13.82 g/cm^{3}$ \cite{brixner1963tage2, yuan2015ta5ge3} for the binary solution, this amorphous film exhibits a $\sim$4\% increase in density. This increase in density is attributed to the amorphous nature of the film, which has been seen previously for pure germanium \cite{blanco1986agedensity}.

\begin{figure*}[h!]
    \centering
    \includegraphics[width=\linewidth]{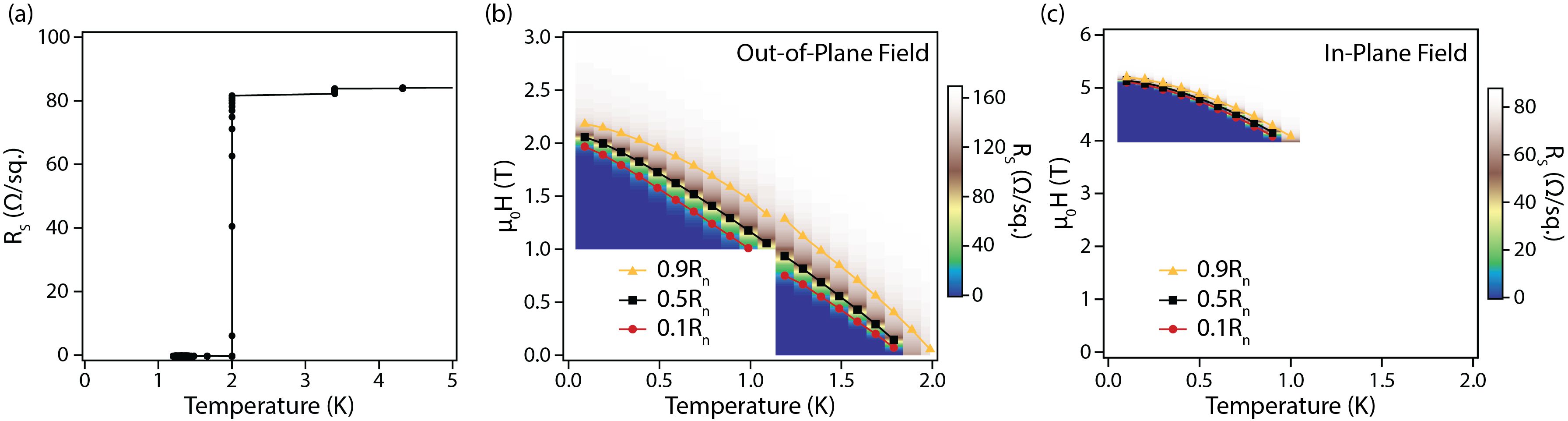}
    \caption{Magnetotransport data of tantalum germanide film. \textbf{(a)} Cooldown trace of sheet resistance, R$_{S}$. We see a zero-resistance transition near 2K.
    \textbf{(b)} R$_{S}$ map as a function of applied out-of-plane field and temperature. Lines are fits to the 0.9, 0.5, and 0.1R$_{n}$ values. \textbf{(c)} R$_{S}$ map for in-plane magnetic field configuration.}
    \label{transport}
\end{figure*}

Magnetotransport measurements of this material are done in an Oxford Triton dry dilution refridgerator with a base temperature of 15mK and a single direction magnet with a maximum field of 14T. Samples are cleaved down to roughly 5x5mm chips. Contacts on the chip are made by annealing In-Sn eutectic into the four corners in a Van der Pauw geometry which are then contacted to the sample board via gold wires. The results of these measurements are presented in Figure \ref{transport}. We see a sharp transition to a zero resistance state at roughly 2K upon cooldown, as shown in Figure \ref{transport}a. To investigate the field dependance, we sweep out-of-plane and in-plane magnetic field as shown in Figs. \ref{transport}b and c, respectively. The resistance surface presented in Fig. \ref{transport}b shows a very sharp transition from the superconducting phase to the normal state, suggestive of type 1 superconductivity. This is observed in the tight spread of 0.9R$_{n}$, 0.5R$_{n}$, and 0.1R$_{n}$ contours overlaid on the color map. For out-of-plane field configuration we observe a critical applied field of $H_{C} \sim 1.88T$. The critical field fitting gives us a critical temperature of roughly 1.8K. This temperature is similar to the reported range of $T_{C}$ of 2.3K-2.75K for compositions of $x$=17-55\% in the Ta$_{x}$Ge$_{1-x}$ alloy system \cite{ghosh1977tmgereview}, which also agrees well with our RBS composition data. With an in-plane field configuration we measure a critical applied field of $H_{C} \sim 5.1T$ that also exhibits a sharp superconducting transition up to 1K. We present this data in Fig. \ref{transport}c in which we overlay the 0.9R$_{n}$, 0.5R$_{n}$, and 0.1R$_{n}$ contours. The variation in sheet resistance between Fig. \ref{transport}b and Figs. \ref{transport}a and c are a result of oxidation in the film over the course of roughly six months causing an increase in film resistance. The measurements were otherwise conducted on the exact same sample.

In conclusion, we presented here the growth, structure, and transport characteristics of diffusion-grown tantalum germanide thin films grown via MBE. The tantalum germanide films are formed by a semi-rate-limited diffusion process in which the e-beam evaporated Ta diffuses into the surface that preserves a $3:2$ Ta:Ge composition ratio within the film region. These films form a small native oxide layer with sharp interfaces and exhibit a roughly 4$\%$ increase in material density. Transport shows a reasonable critical temperature of $\sim$1.8-2K, but more we find a critical out-of-plane and in-plane critical field of $H_{C}^{\perp} \sim 1.88T$ and $H_{C}^{\parallel} \sim 5.1T$, respectively. With such a high critical field, ease of semiconductor integration, and the preservation of sharp S-Sm interfaces, this material shows promise for integration into cQED devices. 

\section{Supplemental Material}

The supplemental material contains additional details on the EDS and XPS experimental considerations and analysis. The data presented herein shows the depth-resolved elemental profiles for the tantalum germanide film as an image via EDS as well as XPS core level measurements as a function of sputtering depth.

\begin{acknowledgments}
The authors would like to acknowledge funding support for this project by AFOSR award FA9550-21-1-0338. 
We gratefully acknowledge the use of facilities and instrumentation supported by NSF through the Cornell University Materials Research Science and Engineering Center DMR-1719875.
\end{acknowledgments}



\section*{Data Availability Statement}

The data that support the findings of this study are available from the corresponding author upon reasonable request.

\bibliography{bibliography}

\end{document}


\preprint{AIP/123-QED}

\title{Supplemental Material for Molecular beam epitaxy growth of superconducting tantalum germanide}
\author{Patrick J. Strohbeen}
\affiliation{ 
Center for Quantum Information Physics, Department of Physics, New York University, New York, NY 10003 USA
}%
\author{Tathagata Banerjee}
\affiliation{
School of Applied and Engineering Physics, Cornell University, Ithaca, NY 14853 USA
}%
\author{Aurelia M. Brook}
\author{Ido Levy}
\affiliation{ 
Center for Quantum Information Physics, Department of Physics, New York University, New York, NY 10003 USA
}%
\author{Wendy L. Sarney}
\affiliation{%
Army Research Directorate, DEVCOM Army Research Laboratory, Adelphi, MD 20783 USA
}%
\author{Jechiel van Dijk}
\author{Hayden Orth}
\author{Melissa Mikalsen}
\author{Valla Fatemi}
\affiliation{
School of Applied and Engineering Physics, Cornell University, Ithaca, NY 14853 USA
}%
\author{Javad Shabani}
\thanks{corresponding author: jshabani@nyu.edu}
\affiliation{ 
Center for Quantum Information Physics, Department of Physics, New York University, New York, NY 10003 USA
}%

\date{\today}

\maketitle

\begin{figure}[h!]
    \centering
    \includegraphics[width=0.8\linewidth]{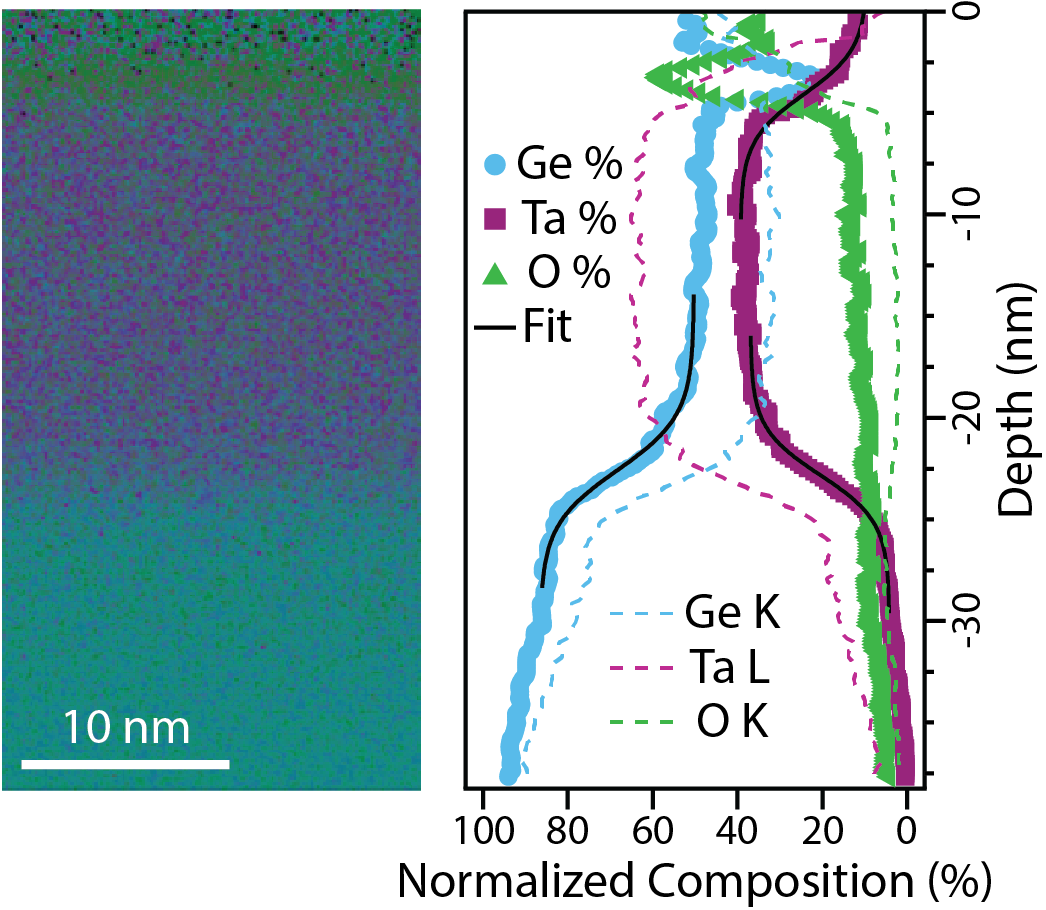}
    \caption{EDS measurement of the film and native oxide regions. The datapoints in the right-hand plot are averages across the entire frame on the left-hand side and the overlaid black line traces are the sigmoid fits to the composition edge profiles. The dashed lines are the normalized X-ray emission line signals at the detector.}
    \label{s1-eds}
\end{figure}

\textbf{Energy Dispersive X-ray Spectroscopy.} Energy dispersive X-ray spectroscopy (EDS) data is collected using an Oxford Aztec system and is presented in \ref{s1-eds} in both the normalized atomic concentration and normalized Ge K-, Ta L-, and O K-edge signals. We fit the EDS profiles to a sigmoid function \cite{dyck2017aptsigmoid}, the form of which is presented in Eq. \ref{sigmoid}, where $C$ is a scaling parameter that defines the maximum concentration, $x_{0}$ is the midpoint of the function, $V$ is the baseline concentration, and $\tau$ is the rate of change across the interface. The results of these fits are presented in Table \ref{sig_fit}, where we report the ``Max'' value as the summation of the base value and the $C$ parameter from the fit.

\begin{equation}
    f(x) = \frac{C}{1+{\rm exp}(\frac{x_{0} \pm x}{\tau})}+V
    \label{sigmoid}
\end{equation}

\begin{table}[h!]
    \centering
    \begin{tabular}{||c|c|c|c|c||}
        \hline
        Layer & \thead{Base, $V$ \\ (At.$\%$} & \thead{Max, (Base + $C$) \\ (At.$\%$)} & \thead{Rate, $\tau$ \\ (At.$\%$ /nm)} & \thead{x$_{0}$ \\ (nm)} \\
        \hline\hline
        \makecell{Germanium \\ Backside} & 0.86 & 0.50 & -1.33 & -22.8 \\
        \hline
        \makecell{Ta$_{x}$Ge$_{1-x}$ \\ Surface} & 0.10 & 0.40 & 1.12 & -4.5 \\
        \hline
        \makecell{Ta$_{x}$Ge$_{1-x}$ \\ Backside} & 0.05 & 0.37 & 1.13 & -22.6 \\ 
        \hline
    \end{tabular}
    \caption{EDS sigmoid fitting results.}
    \label{sig_fit}
\end{table}

Using the sigmoid fitting results, we define the film thickness as the value of $x_{0}^{surf} - x_{0}^{back}$. From this, we measure a thickness of $\sim$18.2~nm for the film, where we define $x_{0}^{back}$ as the average value between the germanium and tantalum germanide fits,which agrees well with our value extracted from the TEM imaging. We define the composition of the Ta$_{x}$Ge$_{1-x}$ region to be the average tantalum composition across the film region, $\sim$38.5$\%$. Taking this value, we estimate the tantalum germanide phase to have a $Ta:Ge$ ratio of $\sim 38.5:61.5$. From this composition we approximate the stoichiometry of the film to be $\sim$Ta$_{2}$Ge$_{3}$. From the EDS we also estimate the native oxide to have a $O:Ta:Ge$ ratio of $\sim 56:22:22$. From this, we approximate the stoichiometry of the oxide region to be $\sim$Ta$_{2}$Ge$_{2}$O$_{9}$, or a solution of one part Ta$_{2}$O$_{5}$ and two parts GeO$_{2}$. However, EDS is well known to  underestimate elemental species, which we attribute to inaccuracies such as oxidation during sample transfer, lineshape overlaps, and not having a well-defined standard for this material \cite{gauvin2012edserrors}. Furthermore, experimental considerations such as insufficient dead-time at the detector can cause high energy X-rays to wash out the signal from the low energy X-rays, which has been observed previously in oxide materials \cite{sarney2023hzo}. This effect is observed in the large discrepancy between the measured signal at the detector (dashed lines in Fig. \ref{s1-eds}) and the resulting At. \%. Therefore, we use the EDS profiles presented here exclusively to report spacial homogeneity of the Ta species as well as define the thickness of the film region.

\begin{figure}[h!]
    \centering
    \includegraphics[width=0.8\linewidth]{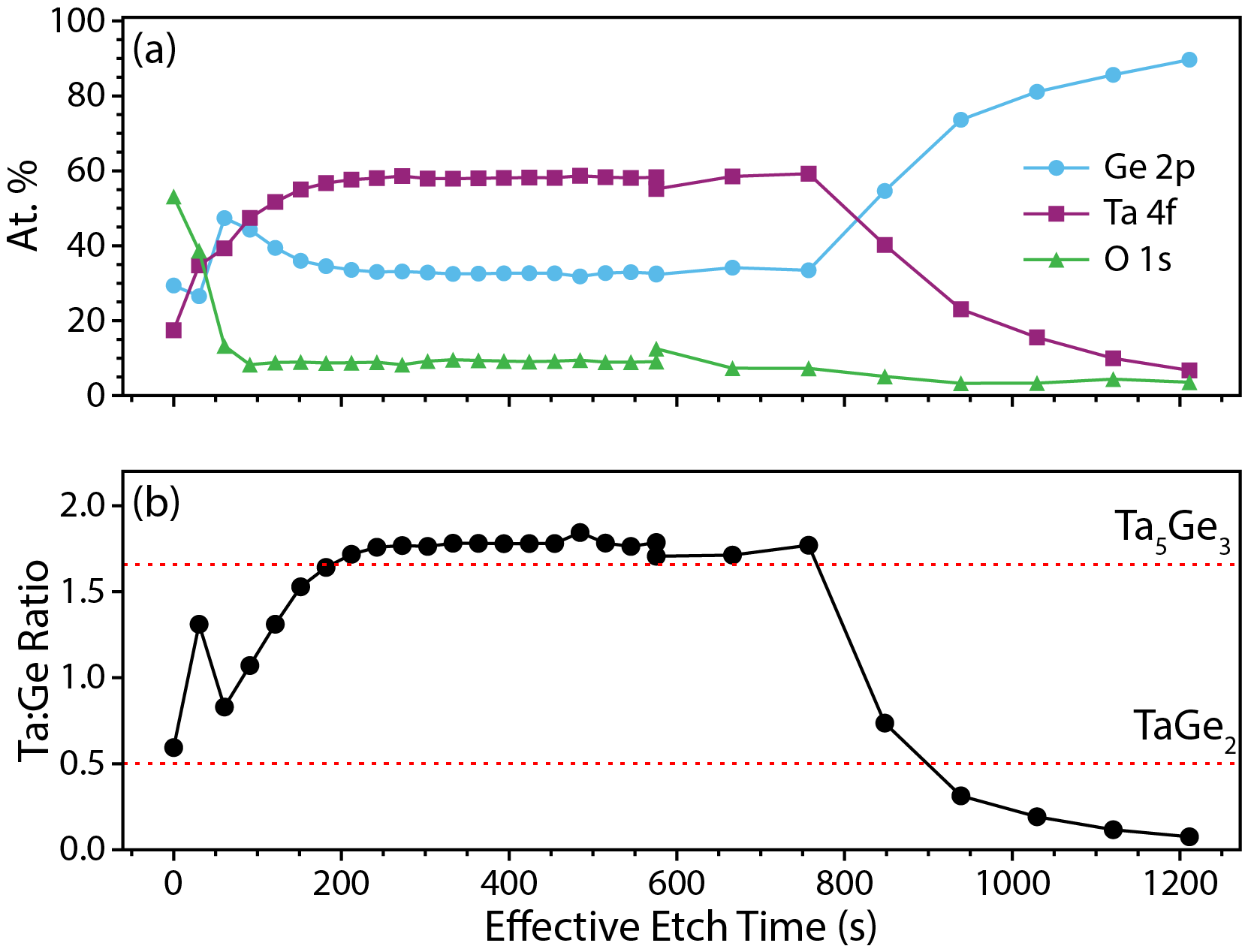}
    \caption{\textbf{(a)} Atomic concentration of Ta, Ge, and O core levels as a function of effective etch time. \textbf{(b)} Ta:Ge ratio. Dashed lines indicate the ratios for Ta$_{5}$Ge$_{3}$ and TaGe$_{2}$. We note the nominally constant composition across the film.}
    \label{s2-xps}
\end{figure}

\textbf{X-ray Photoemission Spectroscopy.} Depth profiling was done with X-ray photoemission spectroscopy (XPS) in a Thermo Nexsa G2 Surface Analysis System with a chamber pressure of $\sim 3.2 \times 10^{-7}$~mBar. The excitation used for these measurements were a monochromated Al K$\alpha$ x-ray source (h$\nu$ = 1486.6~eV) focused to a spot size of 400~$\mu$m. The measurements were conducted with an energy resolution of 0.4~eV and the Fermi level was calibrated to a silver standard inside the analysis chamber. Etching was conducted using monoatomic Ar ions at a constant etch rate until 575~s, at which point the etch rate was tripled. This tripling of etch rate is presented as a tripling of etch time in Fig. \ref{s2-xps}. Cycling time between etches was kept below 5 minutes to limit re-oxidation of the surface. Core level fitting was done assuming a Voigt lineshape using CasaXPS with a Shirley background subtraction. To extract atomic percentages, core levels are normalized to the integrated intensity of the O 1s core level.

\textbf{Temperature Dependent Measurement.} Data was recorded every 5s while the thermometer only updates every 2 minutes, resulting in clusters of 23 data points at each temperature. The material undergoes a superconducting transition partway during this time period, as such we are reporting the temperature at which the transition starts.

\bibliography{bibliography}